# Evaluation of scatter rejection and correction performance of 2D antiscatter grids in cone beam computed tomography

Yeonok Park, Timur Alexeev, Brian Miller, Moyed Miften, Cem Altunbas[*]

Department of Radiation Oncology
University of Colorado School of Medicine
1665 Aurora Court, Suite 1032, Mail stop F-706 Aurora, CO 80045

[*] Email: cem.altunbas@cuanschutz.edu

**Purpose:** We have been investigating 2D antiscatter grids (2D ASG) to reduce scatter fluence and improve image quality in cone beam computed tomography (CBCT). In this work, two different aspects of 2D ASGs, their scatter rejection and correction capability, were investigated in CBCT experiments. To correct residual scatter transmitted through the 2D ASG, it was used as a scatter measurement device with a novel method: grid-based scatter sampling.

**Methods:** Three focused 2D ASG prototypes with grid ratios of 8, 12, and 16 were developed for linac-mounted offset detector CBCT geometry. In the first phase, 2D ASGs were used as a scatter rejection device, and the effect of grid ratio on CT number accuracy and contrast-to-noise ratio (CNR) evaluated in CBCT images. In the second phase, the grid-based scatter sampling method was implemented. It exploited the signal modulation characteristics of the 2D ASG's septal shadows to measure and correct residual scatter transmitted through the grid. To evaluate CT number accuracy, the percent change in CT numbers was measured by changing the phantom size from head to pelvis configuration.

**Results:** When 2D ASG was used as a scatter rejection device, CT number accuracy increased and the CT number variation due to change in phantom dimensions was reduced from 23% to 2–6%. A grid ratio of 16 yielded the lowest CT number variation. All three 2D ASGs yielded improvement in CNR, up to a factor of two in pelvis-sized phantoms. When 2D ASG prototypes were used for both scatter rejection and correction, CT number variations were reduced further, to 1.3–2.6%. In comparisons with a clinical CBCT system and a high-performance radiographic ASG, 2D ASG provided higher CT number accuracy under the same imaging conditions.

**Conclusions:** When 2D ASG was used solely as a scatter rejection device, substantial improvement in CT number accuracy could be achieved by increasing the grid ratio. 2D ASGs could also provide significant CNR improvement even at lower grid ratios. When 2D ASG was used in conjunction with the grid-based scatter sampling method, it provided further improvement in CT number accuracy, irrespective of the grid ratio, while preserving 2D ASG's capacity to improve CNR. The combined effect of scatter rejection and residual scatter correction by 2D ASG may accelerate implementation of new techniques in CBCT that require high quantitative accuracy, such as radiotherapy dose calculation and dual energy CBCT.

## 1. Introduction

Scattered radiation is a major cause of image quality degradation in flat-panel detector-(FPD) based cone beam computed tomography (CBCT). Numerous solutions have been proposed to address the scatter problem, which can be divided into three major groups.

The first group is scatter correction methods, where scatter intensity is estimated in projections, and subtracted to obtain a primary-only intensity distribution[1,2]. This approach has been shown to improve CT number accuracy and reduce artifacts in CBCT images. However, this approach has two shortcomings. First, actual imaging conditions—specifically the imaged object—may not be accurately modeled, and hence, scatter generation and estimation in projections may not provide sufficient accuracy. Recent approaches based on Deep Learning scatter estimation, 3D model-based scatter estimation, and hardware-based scatter estimation aim to address this problem[3-7]. Second, scatter correction methods cannot address contrast-to-noise ratio (CNR) degradation due to scatter, because such methods correct mean scatter intensity after the injection of scatter into projections, and stochastic noise due to scatter is not mitigated.

The second group is image restoration methods. These methods do not directly estimate and correct scatter, but they aim to correct degradation of CT number accuracy and CNR in CBCT images. Earlier image restoration methods have used heuristic approaches[8-10], and more recently, Deep

Learning-based image synthesis methods have been explored[11,12]. The drawback of the latter is that they require extensive training data sets to generate CBCT images with image quality similar to gold standard multi-detector CT (MDCT).

The third and final group is hardware-based scatter rejection methods, such as radiographic 1D ASGs[13,14]. This approach has the potential to improve both CNR and CT number accuracy, since it suppresses scattered x-rays before they reach the image receptor. Though scatter rejection is, theoretically, the ideal solution to the scatter problem, radiographic ASGs provide only moderate improvement in image quality due to the relatively low efficiency of scatter rejection and the attenuation of primary x-rays by the ASG[15-17].

The recent introduction of the 2D ASG concept represents a new scatter rejection solution with the potential to provide superior CT number accuracy and CNR compared to radiographic antiscatter grids[18-20]. While a previous study on 2D ASGs has demonstrated their utility in CBCT image quality improvement[19], this study has employed relatively low grid ratio, a key ASG parameter affecting the scatter rejection performance, and the boundaries of CBCT image quality improvement by 2D ASGs with higher grid ratios have not been investigated. Moreover, an ASG cannot completely suppress scatter fluence incident on the detector; based on the prior scatter transmission evaluations, 2–8% of scatter fluence is still transmitted through the 2D ASG [21], which is expected to degrade CBCT image quality.

Thus, the aim of this work is twofold: First, to develop and evaluate multiple 2D ASG prototypes with grid ratios of up to 16 in terms of the effect of grid ratio on CT number accuracy and CNR. Second, to implement a new residual scatter correction method in conjunction with 2D ASGs, and investigate its effect on further improvement of CBCT image quality. The new scatter correction method, referred as grid-based scatter sampling (GSS), utilizes a 2D ASG to measure and correct residual scatter intensity[22]. In essence, this approach expands the role of 2D ASG from that of a *scatter rejection* device to a *scatter measurement and correction* device. Proof-of-concept work on the GSS method[22] indicated that further improvement in CT number accuracy could be achieved by correcting residual scatter. However, the utility of this method under different 2D ASG grid ratios has not been investigated. The grid ratio affects both intensity and spatial distribution of residual scatter in projections. If successful residual scatter correction vias GSS can be achieved at lower grid ratios, fabrication of 2D ASGs will be less challenging, and source-ASG alignment[21] will be less of a concern during clinical implementation. Thus, in the current work, the utility of residual scatter correction with GSS method was evaluated as a function of the grid ratio of 2D ASGs.

## 2. Materials and Methods
### 2.1. Grid configurations evaluated

Three 2D ASG prototypes were developed. They were fabricated additively from pure tungsten using a powder bed laser melting process (Philips, Best Netherlands). The septa of the 2D ASGs were focused in an offset detector geometry of a TrueBeam CBCT system (Varian, Palo Alto, CA). The fabricated 2D ASGs had grid ratios of 8, 12, and 16, with respective grid pitches of 2.9, 2, and 2 mm. Such gird pitches aim to improve primary transmission. All 2D ASGs had 0.1 mm septal thickness. These ASGs were referred as R8P3, R12P2, and R16P2.

R12P2 and R16P2 ASGs were 3 cm wide in the cranio-caudal direction and 40 cm long in the lateral direction (Fig. 1), while R8P3 was 2 x 40 cm$^2$. They were coupled to the FPD with an aluminum support frame. While grid dimensions in the cranio-caudal direction were smaller than the detector size, the x-ray exposure field of view covered the full active area of the detector, emulating the scatter conditions of a clinical CBCT scan.

In addition to 2D ASG prototypes, a high-performance focused radiographic grid with a grid ratio of 21 (Smit Rontgen by Philips, Best, Netherlands) was evaluated. The radiographic ASG had fiber interspacers and 1D lead septa.

**Table 1.** Grid configurations used in the study

| Grid name | Grid ratio | Grid pitch (mm) | Septal thickness (mm) | Septum material | Interspace material |
|---|---|---|---|---|---|
| 1D R21 | 21 | 0.28 | 0.036 | Lead | Fiber |
| 2D R8P3 | 8 | 2.91 | 0.1 | Tungsten | Air |
| 2D R12P2 | 12 | 2 | 0.1 | Tungsten | Air |
| 2D R16P2 | 16 | 2 | 0.1 | Tungsten | Air |

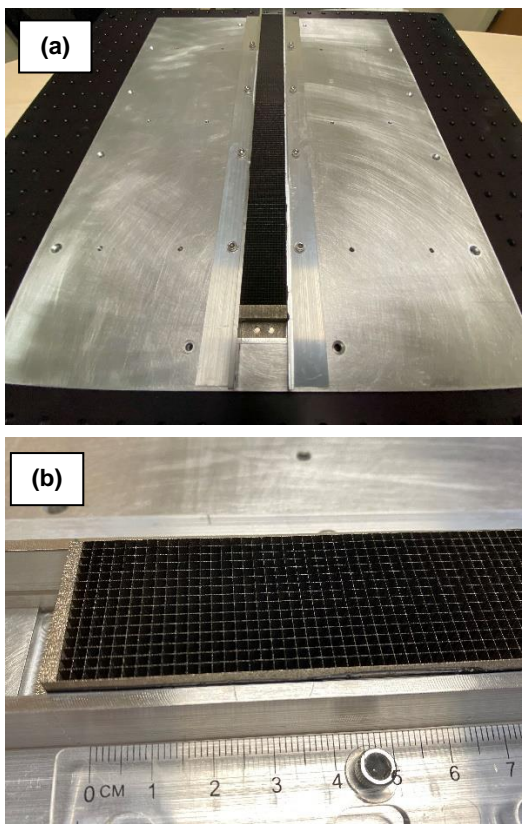

**Fig. 1. (a)** R12P2 ASG mounted on the aluminum support plate. **(b)** Close-up view of the R12P2.

**2.2 Experiment setup and image quality evaluations**

To evaluate the ASGs in Table 1, TrueBeam's standard radiographic ASG was removed and the ASGs under investigation mounted directly on the FPD. CBCT scans of phantoms were acquired in TrueBeam's half-fan, offset detector (large field of view), CBCT geometry (Fig. 2). In addition to phantom scans, flood field scans (i.e., with no object in the beam) were acquired for flat-field correction, and for residual scatter estimation, as described in Section 2.3. The radiation field of view was 40 x 30 cm$^2$ at the detector plane, covering the full active area of the FPD. The source to detector distance was 150 cm, and the FPD's center was offset 16 cm laterally from the piercing point. For inter-grid comparisons, 900 projections (pixel size: 0.388 x 0.388 mm$^2$) were acquired per scan at 125 kVp, 38 mA and 13 ms. A 0.9 mm titanium filter was in place, and the bow tie filter was removed to avoid the scatter and primary intensity modulation it would introduce. For image quality evaluations, the Catphan 504 phantom (Phantom Laboratory, NY) was used in small and large configurations. The small phantom was the Catphan module itself (20 cm diameter cylinder), and the large phantom was a 30 cm by 38 cm elliptic annulus placed around the Catphan.

In each CBCT scan, 900 projections were acquired, and each projection was binned (3 x 3). After flat field correction, images were reconstructed using the FDK filtered backprojection with offset detector weights[19,23-25]. The Hann filter was used and the cutoff frequency set to the Nyquist frequency. Images were reconstructed at a voxel size of 0.9 x 0.9 x 1.0 mm$^3$.

Since 2D ASGs do not completely suppress scatter fluence incident on the detector, residual scatter transmitted through the 2D ASG can still degrade CT number accuracy. To evaluate the effect of residual scatter correction on image quality, CBCT scans with 2D ASGs were corrected using the GSS method[22]. As described in Section 2.3, the GSS method employs the 2D ASG's septal shadows to measure and correct residual scatter in projections, and thus, it was not used for the ASG-free and 1D ASG configurations. With the exception of residual scatter correction, no other post-processing steps (such as beam hardening or image lag correction) were applied to the 2D ASG CBCT scans.

The performance of 2D ASG and GSS method was also compared to a clinical CBCT protocol in Varian TrueBeam. Scans of an electron density phantom (Sun Nuclear Corp., Melbourne, FL), and of standard- and large-sized pelvis phantoms were acquired using the TrueBeam's pelvis protocol, which employed the geometry in Fig. 2. Projections were acquired with a bow tie filter using 125 kVp, 60 mA and 20 msec pulse per projection. Scans with R12P2 ASG were acquired using the same scan parameters, and the residual scatter in projections were corrected using the GSS as described in Section 2.3. The clinical TrueBeam CBCT protocol employs a radiographic 1D ASG with a grid ratio of 10, and performs scatter correction using pencil beam scatter kernel deconvolution[26,27]. In addition, beam hardening correction was implemented. Image reconstruction was performed using the FDK method. While both clinical TrueBeam and 2D ASG scans were acquired using the same imaging dose, image reconstruction and processing parameters (such as reconstruction filter type and cutoff frequency) were not matched between the two. Since these parameters directly affect image noise and may bias CNR values, comparisons between the clinical and 2D ASG CBCT images were focused on evaluations of CT number accuracy.

The same phantoms were also scanned with a Philips Big Bore helical CT for reference purposes. The helical CT was operated at in a 16 x 1.5 mm detector configuration at 120 kVp and reconstructed at 3 mm slice thickness using a body scanning protocol. Since phantom positioning (such as phantom pitch and roll) was slightly different in each image set acquired with the three systems described above, images were fused using rigid image registration before image quality analysis.

To evaluate the improvement in CNR, its value was first calculated using the five contrast objects in the Catphan phantom (Fig. 3(a)). Regions of interests (ROIs) were selected in each contrast object and the adjacent background[21]. The CNR improvement provided by each grid configuration was evaluated relative to ASG-free CBCT images using the CNR improvement factor,

$$K_{CNR} = \frac{CNR_{ASG}}{CNR_{NOASG}} \quad (1)$$

CT number accuracy was evaluated using three different methods. First, the change in CT numbers for a given ROI was evaluated as a function of ASG configuration. An increase in CT numbers for a given material is an indicator of the scatter suppression performance of the ASG. This evaluation was performed using the Teflon insert, ROI #1 in Fig. 3. Teflon is the highest attenuating material in the Catphan phantom, and therefore, the scatter-to-primary ratio in its projections, and the effect of scatter on CT number accuracy, were the largest among all Catphan material types. This analysis was repeated using the phantom's uniform background material, which has Hounsfield Unit (HU) values closer to water.

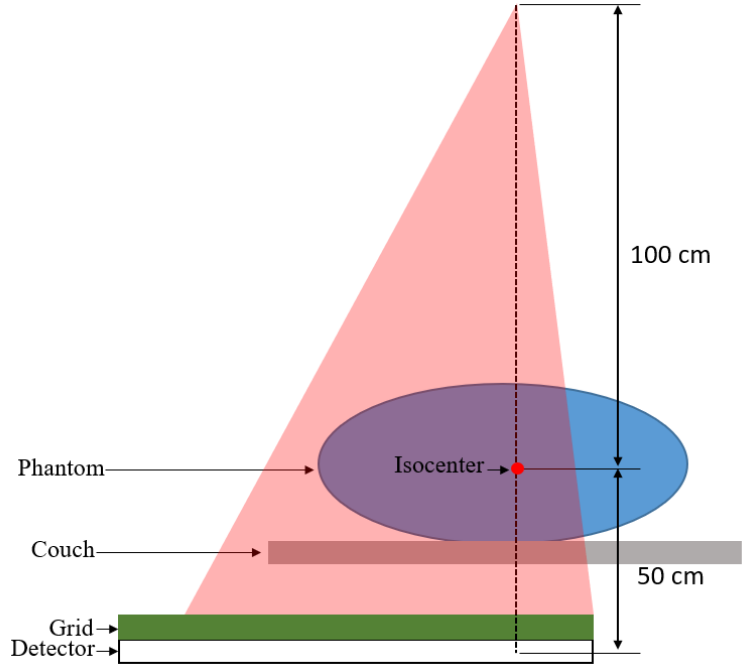

**Fig. 2.** Schematic illustration of the Varian TrueBeam's offset detector CBCT geometry used in experiments.

Second, CT number accuracy was evaluated by measuring the variation in HU values by changing the phantom size from small to large. This is a critical test in CBCT imaging, because HU values may vary drastically as a function of object size. The percent change in HU values for a given material type in small and large phantoms is

$$K_{HU} = 100 \times \frac{|ROI\_HU_{small} - ROI_{HU\,large}|}{ROI\_HU_{small} + 1000} \quad (2)$$

where $K_{HU}$ is the HU loss fraction. Ideally, $K_{HU}$ should be zero, implying that HU value should not change as a function of phantom size.

Finally, HU nonuniformity was evaluated in the uniform background phantom material, which indicated the variation in HU as a function of location in a phantom. 3-mm-wide ring-shaped background ROIs were selected in the phantom body surrounding the material inserts (such as the ROIs indicated in Fig. 12). Mean HU was calculated in each ROI, and the difference between the maximum and minimum of mean HUs among all ROIs was used as the HU nonuniformity metric.

**2.3. Overview of grid-based scatter sampling**

The GSS method is described briefly here for completeness. A detailed description of the method can be found in another publication[22].

When a 2D ASG is on the FPD, primary x-rays incident on the grid septa are stopped, and the primary signal intensity is reduced in pixels underneath grid septa (or in septal shadows). In contrast to primary x-rays, scattered x-rays transmitted through the grid exhibit a different distribution due to their relatively broad angle of incidence on the detector. When grid septa are sufficiently thin, scattered x-rays continue to reach pixels beneath grid septa and scatter intensity remains locally uniform across in grid shadows and adjacent grid holes. The GSS method exploits this difference between primary and scatter intensity distributions to estimate the residual scatter intensity.

Implementation of the GSS method starts with acquisition of a flood field projection (i.e., with no object in the beam and only primary x-rays). Flood projection characterizes the signal intensity reduction due to ASG's septal shadows, and can be compensated for using pixel specific normalization factors, known as the gain map, *GM*,

$$GM(u,v) = \frac{C}{F(u,v)} \quad (3)$$

where *C* is an arbitrary normalization constant, and *F(u,v)* is the flood projection. In a CBCT scan generated by only primary x-rays, septal shadows in a raw projection, $I^{raw}$, are compensated for using

$$I^{cor}(u,v) = I^{raw}(u,v) \times GM(u,v) \quad (4)$$

If we omit signal intensity variations due to attenuation in the imaged object, signal intensity in septal shadows and in adjacent holes will be equal after *GM* correction. However, when residual scatter is present, as in a realistic CBCT scan, *GM* correction overcompensates for signal intensity in septal shadows compared to adjacent grid holes. This is because residual scatter is an *additive* signal in projections and cannot be corrected using a multiplicative *GM* correction. Such a signal intensity difference between two adjacent pixels, one located in a septal shadow and the other in a grid hole, can be described as

$$d(u_1,v_1) = I^{cor}_{septum}(u_1,v_1) - I^{cor}_{hole}(u_2,v_2) \quad (5)$$

Pixels $(u_1,v_1)$ and $(u_2,v_2)$ are located in a septal shadow and a proximal grid hole, respectively. The magnitude of *d* and residual scatter intensity *S* in septal shadows are correlated, and this correlation is exploited to estimate the residual scatter intensity[22],

$$S_{septum}(u_1,v_1) = \frac{d(u_1,v_1)}{GM_{septum}(u_1,v_1) - GM_{hole}(u_2,v_2)} \quad (6)$$

Once *S* is calculated for all pixels in septal shadows, <*S*> in other pixels, i.e. pixels in grid holes, is obtained via interpolation. Subsequently, primary-only projections, $P^{cor}(u,v)$, are obtained by subtracting the interpolated 2D scatter map:

$$P^{cor}(u,v) = [I^{raw}(u,v) - \langle S(u,v) \rangle] \times GM(u,v) \quad (7)$$

This process is repeated for all projections and followed by image reconstruction.

## 3. Results
### 3.1 Effect of scatter rejection on image quality

Fig. 3 shows CBCT images of small and large Catphan phantoms acquired with all ASG configurations. Without an ASG, HU values were severely underestimated, and HU nonuniformity is visible as cupping and shading artifacts. Artifacts in the large phantom are more severe due to increased scatter fraction. With the use of 2D ASGs, shading and cupping artifacts are substantially reduced, as also evident in the HU profiles (Fig. 4).

The mean HU value of Teflon is shown for all ASG configurations in both small and large phantoms in Fig. 5. HU values increased monotonically as a function of 2D ASG grid ratio. In the small phantom, R8P3 exhibited a 317 HU increase relative to NOASG. For the R12P2 and R16P2 ASGs, the differential increase in HU was 62 and 23 HU, respectively. This observation indicated that an increase in the grid ratio from 12 to 16 had a smaller effect on HU accuracy in the small than in the large phantom configuration.

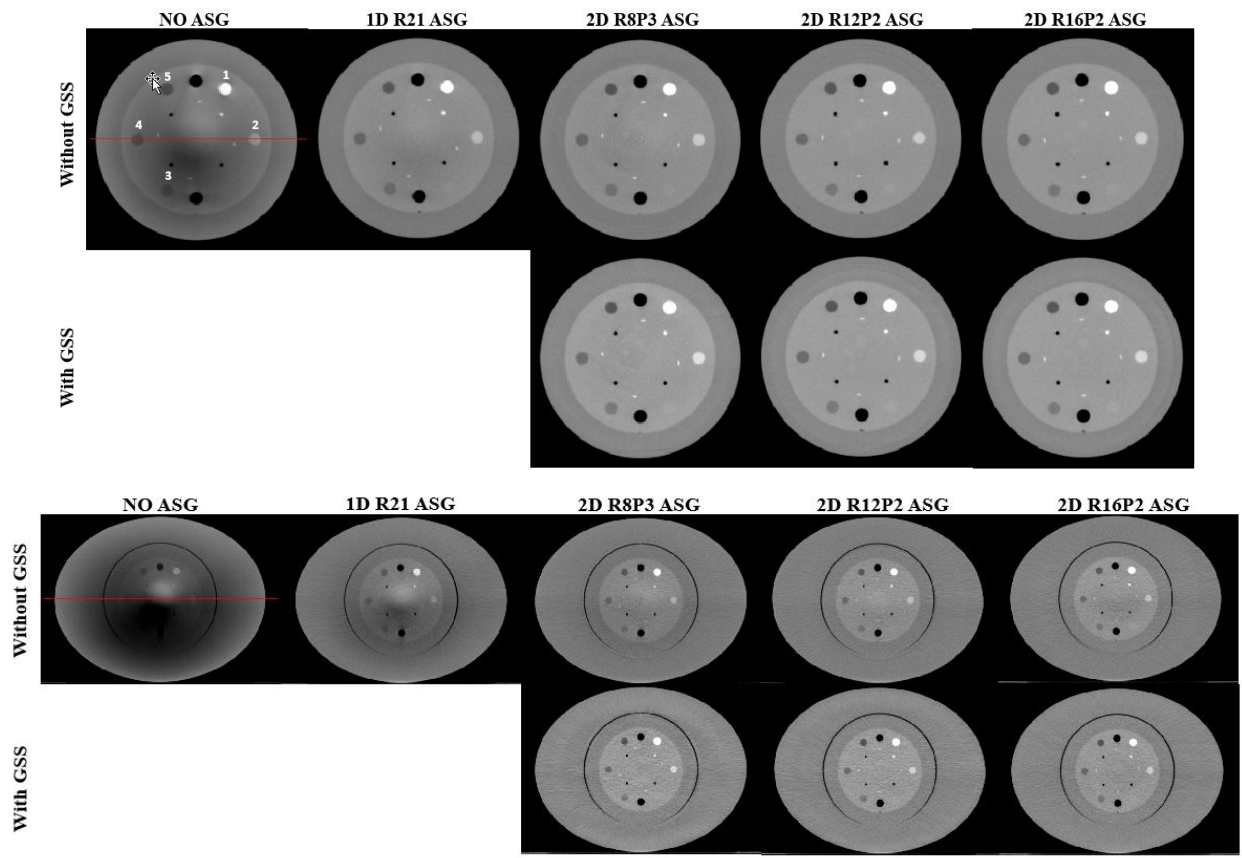

**Fig. 3.** CBCT images of **(a)** head and **(b)** pelvis-sized Catphan phantoms. Top and bottom rows are without and with GSS based residual scatter correction, respectively. HU profiles along the red lines are shown in Fig. 4. HU Window: [-500 500].

While a similar correlation was observed in the large phantom, the increase in HU as a function of grid ratio was larger. R8P3 exhibited a 557 HU increase compared to NOASG. With the R12P2 and R16P2 ASGs, the differential increase was 140 and 109 HU, respectively. Thus, increasing the grid ratio from 12 to 16 had a relatively large impact on HU values in the large phantom configuration.

Similar trends were also observed in the HU analysis of six ROIs in the phantom background material, as shown in Fig. 6. While median HU values increased as a function of grid ratio in both phantoms, the effect of grid ratio was more pronounced in the large phantom.

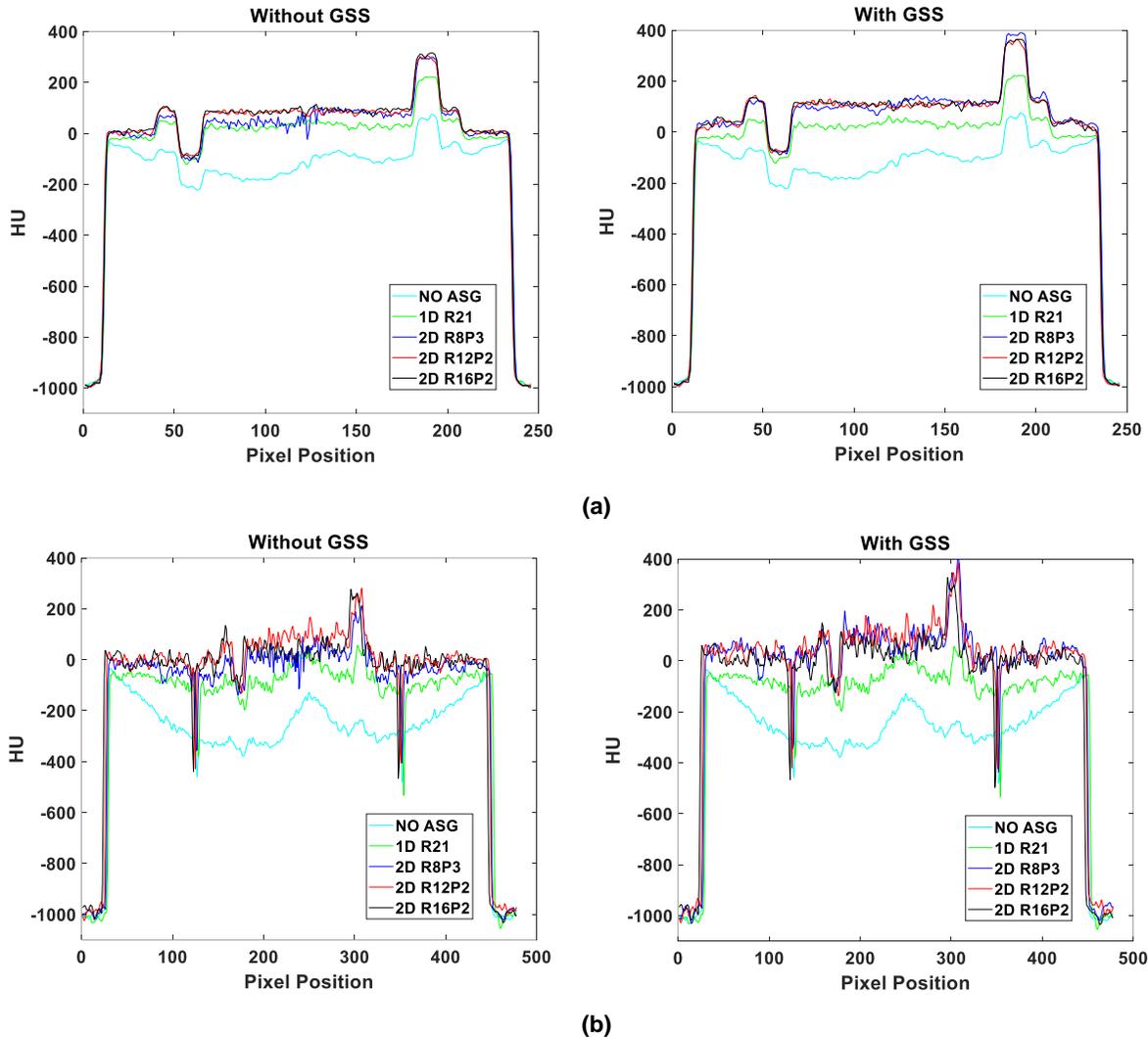

**Fig. 4.** HU profiles along the red line segment indicated in Fig. 3: (a) small and (b) large Catphan phantom; without (left) and with (right) GSS.

While 1D ASG also reduced artifacts and increased CT numbers, the improvement was substantially less compared to all 2D ASGs. For instance, when compared to ASG-free images, 1D ASG increased median HU in the phantom background by 139 and 223 in small and large phantoms, respectively (Fig. 6). Whereas the 2D ASG with the lowest grid ratio (R8P3) increased the median value by 156 and 303 HU under the same conditions.

HU nonuniformity in the phantom background was reduced substantially, as shown by the box and whiskers in Fig. 6. Without an ASG, HU nonuniformity (i.e. the difference between the minimum and maximum whiskers) was 194 HU for the small phantom. For R8P3, the nonuniformity was reduced to 31 HU, whereas nonuniformity was 16 HU for both R12P2 and R16P2, indicating that increasing the grid ratio from 12 to 16 did not impact HU nonuniformity. Likewise, values of HU nonuniformity for R12P2 and R16P2 were comparable in the large phantom (35 and 34 HU, respectively).

To measure the change in CT number accuracy as a function of phantom size, the mean and standard deviation of $K_{HU}$ (again, the percent change in HU values for a given ROI, when phantom size was increased from small to large) was calculated for five material inserts (Fig. 7).

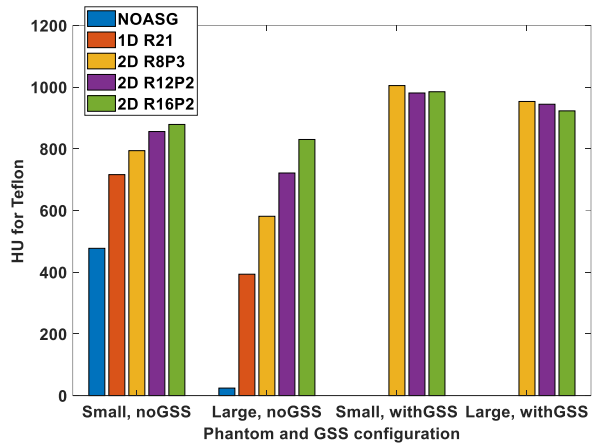

**Fig. 5.** HU values for the Teflon insert in small and large Catphan phantoms. Without the GSS method, HU values increase as a function of 2D ASGs' grid ratio. When phantom size is increased from small to large, a reduction in HU values is noticeable. With the GSS method, HU values among 2D ASGs, and between small and large phantoms, is more consistent.

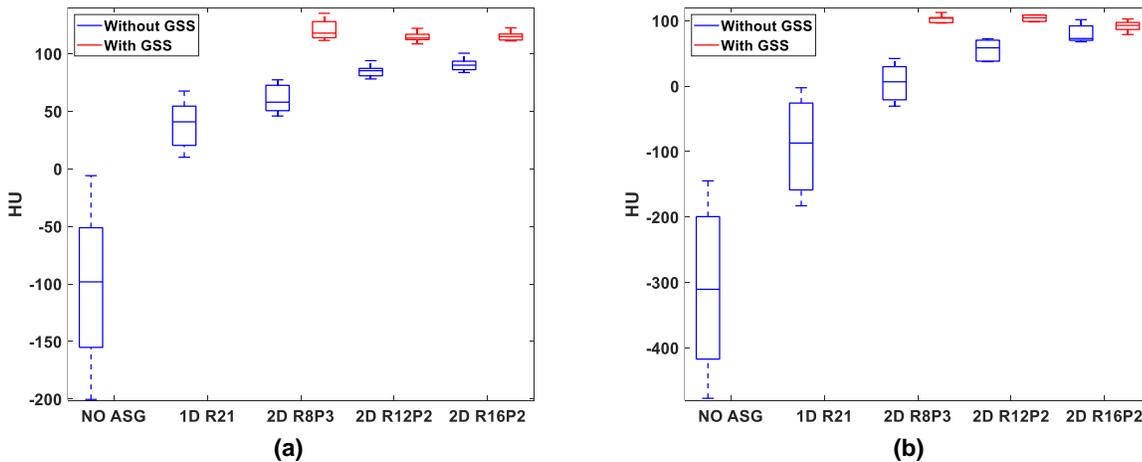

**Fig. 6.** HU statistics of ROIs placed in the phantom background material of **(a)** small and **(b)** large Catphan phantoms. Line, box, and whiskers correspond to median, lower/upper quartiles, and minimum/maximum of mean ROI HU, respectively.

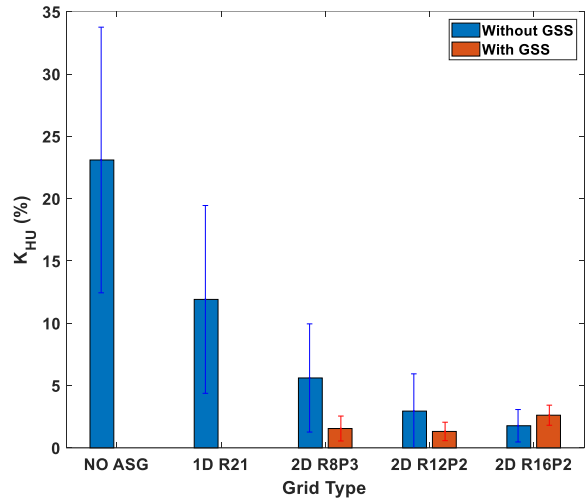

**Fig. 7** Mean and standard deviation of HU loss fraction, $K_{HU}$, for five materials indicated in Fig. 3.

For R8P3, $K_{HU}$ was reduced from 23% to 5.6%, while for R12P2 and R16P2, $K_{HU}$ was reduced further, to 2.9% and 1.8%, respectively, which indicated that CT number inconsistencies due to change in phantom size were reduced by increasing the grid ratio. $K_{HU}$ with 1D ASG was 12%, inferior to all 2D ASGs.

The use of 2D ASGs also improved CNR (Table 2). The CNR improvement factor, $K_{CNR}$, reached up to 1.31 for small and 2.01 for large phantoms. In contrast to the correlation observed between CT number accuracy and 2D ASGs' grid ratio, CNR improvement did not correlate strongly with grid ratio. For example, R12P2 provided the highest $K_{CNR}$ in both small and large phantoms, followed by R16P2 and R8P3. Similar to the trends in CT number accuracy evaluations, a 1D ASG provided lower CNR improvement than any of the 2D ASGs.

**Table 2.** CNR improvement factor, $K_{CNR}$, in small and large Catphan phantoms.

| Phantom size | GSS used? | 1D ASG | 2D ASG | | |
|---|---|---|---|---|---|
| | | R21 | R8P3 | R12P2 | R16P2 |
| Small | No | 1.05 ± 0.09 | 1.09 ± 0.11 | 1.31 ± 0.06 | 1.21 ± 0.15 |
| | Yes | N/A | 1.10 ± 0.12 | 1.31 ± 0.06 | 1.21 ± 0.15 |
| Large | No | 1.49 ± 0.42 | 1.87 ± 0.37 | 2.01 ± 0.38 | 1.95 ± 0.50 |
| | Yes | N/A | 1.86 ± 0.37 | 2.01 ± 0.40 | 2.00 ± 0.50 |

### 3.2 Effect of scatter rejection and residual scatter correction on image quality

The combined effect on CBCT images of scatter rejection by 2D ASG and residual scatter correction using the GSS method is shown in Fig. 3. Since the GSS method requires the use of 2D ASG to correct residual scatter intensity, it was not employed in the 1D ASG and ASG-free configurations.

From a qualitative perspective, further improvement in image quality with residual scatter correction appeared less drastic, because the vast majority of scatter fluence was already rejected by the 2D ASG[21].

However, quantitative analysis indicated that CT numbers increased further with the GSS method, and CT numbers with GSS were more consistent across all 2D ASGs. For example, with residual scatter correction, HU values for Teflon in the small phantom were 1005, 981, and 985 HU for R8P3, R12P2, and R16P2, respectively (Fig. 5), which indicates that the HU value variation for Teflon was only 20 HU among the 2D ASGs, despite the large difference in their grid ratios. A similar variation of 30 HU was observed for the large phantom among the 2D ASGs (953 and 923 HU).

Similar results were obtained in the phantom background material as well (Fig. 6). After residual scatter correction, HU values increased, and similar HU values were obtained for all 2D ASGs. HU nonuniformity was reduced further with residual scatter correction. For example, HU nonuniformity for R8P3 was reduced from 35 to 11 HU in the large phantom. HU nonuniformity values exhibited less variation for all 2D ASGs after residual scatter correction (11–24 HU and 12–24 HU in the small and large phantoms, respectively).

Residual scatter correction also reduced CT number inconsistencies due to change in phantom size (Fig. 7). For R8P3, $K_{HU}$ was reduced from 5.6% to 1.6% after residual correction. $K_{HU}$ for all 2D ASGs were comparable for all 2D ASGs; they were in the range of 1.3–2.6%. All 2D ASGs provided the same or similar $K_{CNR}$ values before and after residual scatter correction (Table 2).

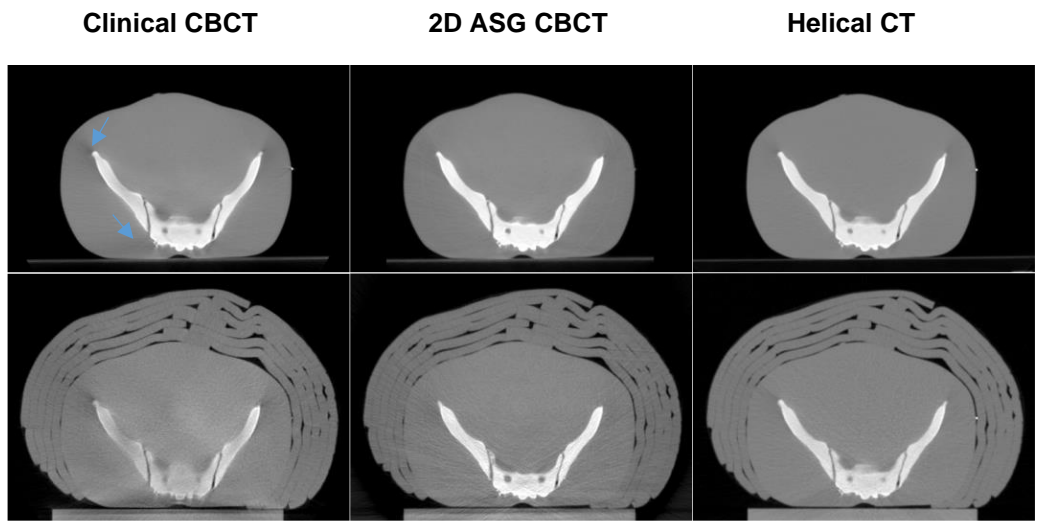

**Fig. 8.** Standard (top) and large pelvis (bottom) phantoms. Clinical CBCT scans were acquired using TrueBeam's standard pelvis CBCT protocol. 2D ASG CBCT scans were acquired with the R12P2 ASG and corrected using the GSS method. Helical CT scans were acquired using a Philips Big Bore 16-slice CT scanner. Blue arrows point to more emphasized shading artifacts in the clinical CBCT image, when compared to 2D ASG CBCT. HU Window: [-1000 1000].

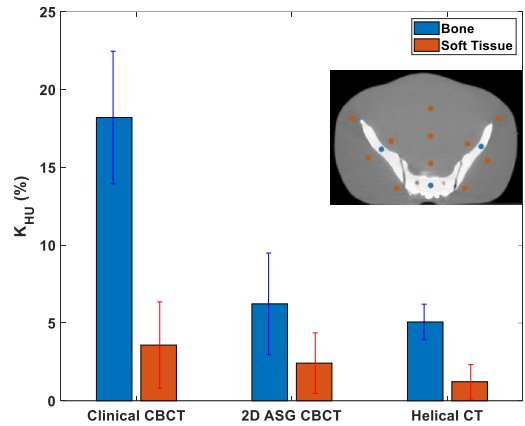

**Fig. 9.** $K_{HU}$ for bone and soft tissue equivalent regions in images of standard and large pelvis phantoms. Bone and soft tissue ROI locations for $K_{HU}$ are shown in the inset.

Only the R12P2 2D ASG was used in comparisons of 2D ASG and the GSS method with TrueBeam's clinical CBCT protocol, because all 2D ASGs provided comparable CT number accuracy when used with the GSS method. Images of standard and large pelvis phantoms are shown in Fig. 8. In visual evaluations, all three systems provided similar HU uniformity in the standard-size phantom. However, CT number degradation was apparent, particularly in the clinical CBCT image, for the large pelvis phantom. Mean $K_{HU}$ values in soft tissue equivalent regions were 3.6%, 2.4%, and 1.2% for clinical CBCT, 2D ASG CBCT, and helical CT images, respectively. HU degradation was drastically higher in bone equivalent regions. Mean $K_{HU}$ values were 18%, 6%, and 5% in Clinical CBCT, 2D ASG CBCT, and helical CT images, respectively.

HU nonuniformities were visible in the clinical CBCT images of the electron density phantom (Fig. 10). Cupping artifacts were introduced in bone equivalent material inserts, which can be seen better in the HU profiles (Fig. 11). The use of 2D ASG and the GSS method reduced such nonuniformities. HU nonuniformity was analyzed in the phantom body (Fig. 12), with fourteen ring-shaped ROIs placed around material inserts, as indicated in the inset. HU nonuniformity, the difference between the min/max whiskers, was 165 HU in clinical CBCT images, but was reduced to 63 HU and 25 HU in 2D ASG and helical CT images, respectively.

**Clinical CBCT**  **2D ASG CBCT**  **Helical CT**

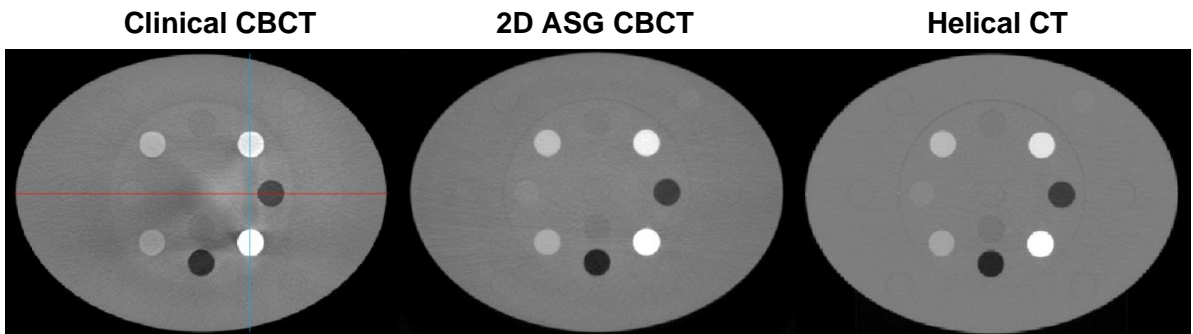

**Fig. 10.** Images of the electron density phantom. Clinical CBCT scan was acquired using TrueBeam's standard pelvis protocol using a bow tie filter. A 2D ASG scan was acquired using the same scan parameters as in the TrueBeam pelvis protocol. HU profiles along the red and blue lines were plotted in Fig. 11. HU window: [-1000 1000].

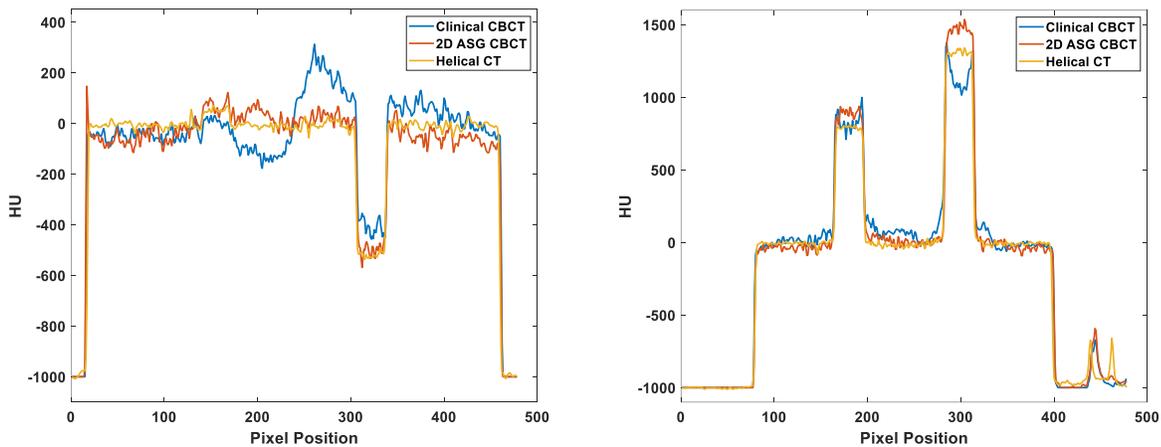

**Fig. 11.** HU profiles along the red (left) and blue (right) line segments indicated in Fig. 10.

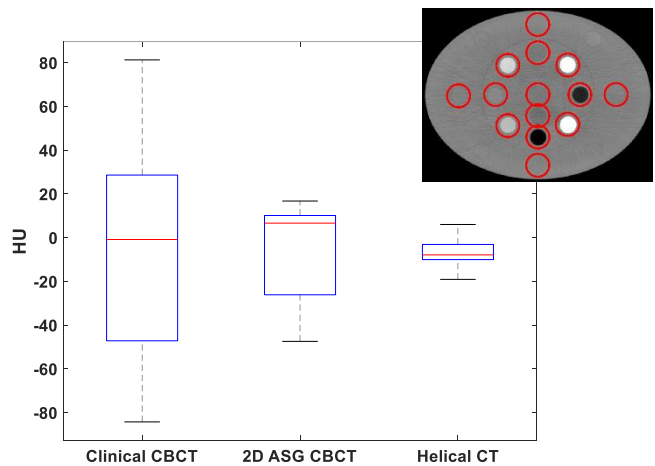

**Fig. 12.** Water HU nonuniformity measured in fourteen water equivalent ROIs in the electron density phantom. ROIs were 3-mm-wide ring structures surrounding the material inserts in the electron density phantom, as indicated in the inset. Line, box, and whiskers represent median, 25th/75th percentiles, and min/max values of mean ROI HU, respectively.

## 4. Discussion

This study evaluated two different functions of 2D ASGs in the context of linac-mounted CBCT imaging: (1) scatter rejection and (2) scatter correction properties.

When 2D ASG was used solely as a scatter rejection device, substantial improvement in CT number accuracy was observed, and as expected, the improvement was proportional to its grid ratio. A major question is how high the 2D ASG's grid ratio should be to achieve high CT number accuracy. While the answer depends on the imaging task, our results support several conclusions. First, the properties of the imaged object, and the scatter fraction in projections, matter in determining the grid ratio. In our study, increasing the grid ratio from 12 to 16 provided small returns in CT number accuracy in the small phantom. Thus, moderate grid ratios of around 12 appear to be sufficient for imaging head-sized objects.

No such trend was observed in the large phantom, which implies that grid ratios higher than 16 might be needed to achieve high CT number accuracy in pelvis-sized objects. While it might be feasible to build ASGs with higher grid ratios using additive manufacturing processes[28], the main challenge lies in the precise alignment of such ASGs toward the focal spot[21]. Imperfect mechanical alignment, or deviation in focal spot position due to gantry sag or x-ray tube characteristics, may cause lower primary transmission through the 2D ASG, increasing image noise and image artifacts.

The need for higher grid ratios can be mitigated by correcting the residual scatter intensity that is not rejected by the 2D ASG. The GSS method investigated in this work achieved residual scatter correction by employing the 2D ASG as a residual scatter measurement device. Once the residual scatter in projections was measured and corrected, the application of GSS enabled even higher CT number accuracy than was possible with the 2D ASG with a grid ratio of 16 alone. In fact, comparable CT number accuracy was observed in a grid ratio range of 8 to16, indicating that the role of grid ratio is minimized if residual scatter is corrected using the GSS method. These results also indicate that the GSS method works reliably under different residual scatter intensity levels, as imposed by phantom dimensions and 2D ASG geometries investigated in this work.

If we again ask how high the grid ratio should be to achieve high CT number accuracy, we can state that the answer depends on whether the 2D ASG is used in conjunction with the GSS method. With the GSS method, a grid ratio of 8, or even lower, may provide sufficient CT number accuracy even for pelvis-sized objects. However, a potential drawback of using lower grid ratios is the increase in ASG-induced artifacts in CBCT images[22], because 2D ASG's septal shadows cannot be suppressed with standard flat-field correction methods[29] due to residual scatter transmitted through the 2D ASG. Lower grid ratios will increase residual scatter intensity in projections. If residual scatter is not fully corrected via the GSS method, it may cause suboptimal suppression of 2D ASG's septal shadows in projections after flat-field correction. Thus, suppression of ASG-induced artifacts should also be considered when selecting the optimal grid ratio.

In addition to improved CT number accuracy, 2D ASGs provided up to two-fold in CNR in pelvis-sized phantoms. The CNR improvement remained same after GSS correction, because this approach reduces the bias in projection signal amplitude due to scatter, and a corresponding improvement in CT number accuracy. Stochastic noise due scatter is not removed, and CNR values remain the same—as is typical for all scatter correction methods [1]. The effect of grid ratio on CNR improvement was small, a result in agreement with prior work on the analysis of 2D ASG projection images [21]. This is the case because residual scatter fraction in projections is low even with lower grid ratio 2D ASGs. For example, a 2D ASG with a grid ratio of 8 reduces the scatter-to-primary ratio (SPR) below ~0.4[18] for water-like phantoms, whereas a grid ratio of 16 reduces the SPR below ~0.2[21]. While such a reduction in SPR has a profound influence on CT number accuracy, as demonstrated in Figs. 5–7, it has a small effect on CNR values[21,30] due to CNR's $1/\sqrt{1+SPR}$ relationship with SPR. The primary transmission of all three 2D ASGs were in the range of 82-86%, and therefore, primary transmission is not expected to have a large influence on the variation of CNR improvement among 2D ASGs.

Another conclusion of this study is that increasing the grid ratio of a radiographic 1D ASG yields inferior improvement in CT number accuracy and CNR compared to using 2D ASGs with lower grid ratios. For example, R8P3, with a grid ratio of 8, provided both higher CT number accuracy and CNR than the radiographic 1D ASG with a grid ratio of 21. While one may consider increasing the grid ratio of 1D ASG beyond 21, the literature suggests this approach is less likely to yield significant improvement in image quality[13,31].

The TrueBeam clinical CBCT system used in this work employs a radiographic 1D ASG and pencil beam scatter kernel deconvolution. The 2D ASG and GSS method provided higher CT number accuracy than the clinical CBCT. Even though helical CT provided better CT number accuracy, our reconstructions with 2D ASG did not employ other data correction methods, such as beam hardening and image lag correction. The former is important in conducting 2D ASG imaging experiments with TrueBeam CBCT because the bow tie filter introduces additional beam hardening at the periphery of the imaging volume. With the implementation of such data correction methods, the CT number accuracy of 2D ASG CBCT may more closely approach that of helical CT. Recent work has developed a more accurate scatter estimation algorithm and iterative reconstruction approach, iCBCT, for TrueBeam CBCT[4]. In the future, we plan to implement a comparable iterative reconstruction algorithm, and compare 2D ASG and the GSS method with the performance of iCBCT.

**5. Conclusions**

A study on the scatter rejection and correction properties of 2D ASGs for CBCT imaging was presented. When 2D ASG was used as a scatter rejection device, it achieved significantly higher CT number accuracy and CNR than a high-performance radiographic ASG. By combining the scatter rejection capability of 2D ASGs with residual scatter correction via GSS method, even higher CT number accuracy was achieved at lower grid ratios. This approach also provided better CT number accuracy than the standard scatter mitigation methods used in linac-mounted CBCT systems.

Improved CT number accuracy with a 2D ASG approach may enable or accelerate implementation of new techniques in CBCT imaging, such as dual energy imaging and the extraction of radiomics features. While such techniques have been increasingly utilized in multi-detector CT imaging, their utilization in CBCT is limited due to poor quantitative accuracy. In radiation therapy, improved CT number accuracy and CNR may yield accurate CBCT-based dose calculations and better soft tissue visualization, which may in turn support the implementation of adaptive radiation therapy concepts in CBCT-guided radiation therapy.

**Acknowledgements**

This work was funded in part by grants from NIH/NCI (R21CA198462) and the Cancer League of Colorado.